\documentclass[final,1p,times]{elsarticle}
\usepackage[usenames]{color}
\usepackage{amssymb}
\usepackage{soul}
\journal{Physica A}
\begin{document}

\begin{frontmatter}

\title{Dynamics of HIV Infection: an entropic-energetic view}

\author[label2]{Ram\'on E. R. Gonz\'alez\corref{cor1}}
\cortext[cor1]{Corresponding author}
\ead{ramayo\_g@yahoo.com.br}
\author[label2]{P. H. Figueir\^edo}
\address[label2]{Laborat\'orio de Sistemas Complexos e Universalidades. Departamento de F\'{\i}sica, Universidade Federal Rural de Pernambuco, 52171-900, Recife, Pernambuco, Brazil.}


\author[label3]{S. Coutinho}
\address[label3]{Laborat\'orio de F\'{\i}sica Te\'orica e Computacional, Departamento de F\'{\i}sica, Universidade Federal de Pernambuco, 50670-901, Recife, Pernambuco, Brazil.}


\begin{abstract}
We propose a time-parameterized analogy between the thermodynamic behavior of a 3-level energy system and the progression of the HIV infection described by the cell population evolution generated by an appropriated cellular automata model. The development of internal energy and its fluctuations, and of the entropy of the 3-level energy system allows the identification of an \emph{effective temperature} that uniquely characterizes the three main stages of the dynamic process of HIV infection (primary infection, clinical latency, and development of AIDS). Furthermore, this \emph{thermodynamical} equivalence allows obtaining the instants associated with particular time intervals of the evolution of internal energy and entropy, which are not quantitatively accessible by the usual dynamic models based on differential equations or cellular automata. The maximum entropy point, which marks the threshold between the states of positive and negative temperatures, also corresponds to the onset of the immune system's 
exhaustion and the concomitant and inexorable progression to AIDS.  Such point lies in the time interval where the energy \emph{inversion} mechanisms in the populations of non-infected and infected cells occur. This time lag is also characterized by considerable fluctuations of the internal energy when different (patient) samples are compared.
\end{abstract}
\begin{keyword}
HIV infection \sep Dynamics of evolution \sep Entropic-energetic description \sep Cellular Automata
\end{keyword}
\end{frontmatter}

\section{Introduction.}\label{sec1}

The dynamics of HIV infection has been intensively investigated through mathematical models based on differential equations \cite{perelson99,nowak00} or by computational simulation of models based on cellular automata (CA)  \cite{kougias90,pandey90,pandey91,haase99,santos01a,bernaschi02}. The objectives of these studies are quantifying and analyzing the underlying dynamic processes, testing hypotheses and assisting in the establishment of treatment strategies and protocols. HIV infection leads to a progressive reduction in the number CD4$^+$ T cells. HIV viral load and CD4 cell counts are the main parameters used to describe the infection process and to define treatment protocols and therapy strategies. The mathematical models generally describe the evolution of viral burden and the CD4$^+$ T cells population in the healthy, infected and dead cells states. 

In this work, we present a ``thermodynamical'' vision for the progression of the macroscopic states of  the course of the HIV infection after the primary infection by means of a CA model designed to describe the dynamics of HIV infection \cite{santos01a}.  We assume that at each time step of the automata evolution (which corresponds to one week) the macroscopic states are viewed as \emph{thermodynamical equilibrium states} with well-defined energy and entropy according to the average populations of the CD4$^+$ T cells in the healthy, infected and dead cells states of the CA model. 

The evolution of healthy (non-infected) and infected CD4$^+$ T cell concentrations are used to characterize the three distinct stages of the dynamics of HIV infection, the  primary infection (PI), the clinical latency (CL), and the AIDS development, which occur in two different time scales. The initial PI phase  is quite similar to that in others in more common viral infections: the viral load reaches its maximum five to six weeks after contamination and endures  of the order of 10 to 12 weeks after which  the viral load decreases to almost undetectable levels. The clinical latency is characterized initially by a very high concentration of healthy cells but gradually decreasing in the time scale of the order of 2 to 10 years or more, leading to impairment of essential functions of the immune system \cite{zdenek2006}. During CL phase, the virus remains \textit{latent} in the body in compartments where they are not identified and eliminated by the immune system. Eventually, such viruses are reactivated leading 
to the development of the acquired immune deficiency syndrome (AIDS).  

Cellular automata (CA) models are spatially structured and therefore more appropriate and efficient to describe local interactions and correlations between cells that participate in the dynamics between the immune response and the viral particles. The model proposed in \cite{santos01a} was specially designed to describe the entire course of HIV infection, from the primary stage to the development of AIDS, by modeling the dynamic behavior of the T-cell population in lymph nodes. In its rules of evolution, this model tests the combination of two important factors: the high viral mutation rate and the spatial location of the infected T cells in the lymph nodes. The results qualitatively reproduce the three stages of the dynamics of HIV infection with their compatible and particular time scales. Such model has inspired other authors to propose new models to investigate other aspects of the infection, including the processes of antiretroviral 
therapies \cite{sloot02,benyoussef03,peer04,solovey04,shi08,figueiredo08b,burkhead09,precharattana11,ramayo12, ramayo13}.
 
In this work, we re-examine the model proposed in \cite{santos01a} from the perspective of the thermodynamic behavior of a three-level energy system. This correspondence highlights important aspects of the dynamics of HIV infection that are not revealed by the usual time series analysis methods. In the following subsection, we describe the CA model and its visualization as a three-level system. In Section \ref{sec2}, we formulate the three-level specific model to HIV dynamics by setting the relevant parameters for the associated energy spectrum. The description of the energy and entropy behaviors related to HIV dynamics are analyzed and broadly discussed in Section 3, and a summary of the findings is presented in Section 4.

\subsection{Cellular automata based models}\label{subsec11}

The CA model proposed in \cite{santos01a} considers the population of CD4$^+$T cells  located in the lymph nodes and the local interactions between its different states, \emph{healthy, infected} (types $A$ e $B$) or \emph{dead}, which evolve differently in the three phases of infection. The spatial structure of the lymph nodes designed simply by a $(L\times L$)-square lattice of size $L = 700$ assuming periodic boundary conditions. The active neighborhood for the interaction between the cells considers the first eight closest neighbors (\emph{Moore neighborhood}) and the initial configuration of the system consists of a fixed fraction of infected cells ($p_{\tiny{HIV}}$) randomly distributed in the network filled with healthy cells. The temporal evolution of the cell states is done synchronously in unit time steps corresponding to one week.

Four simple rules govern the dynamics of the automata, in which some characteristic parameters of the process play an important role. The evolution process of the infection progression occurs by contact of a healthy cell and an infected cell. Healthy cells may become infected, at the next instant of time, if there is at least one infected cell type A and/or $R$ infected cells type B in their active vicinity. Infected cells of type A possess a high capacity of contamination while those of type B have their ability diminished since they have already been identified by the immune system through the specific immune response. The parameter $R$, ($1\leq R\leq 8$) regulates the decrease in the transmission capacity of the infection between the cells and in this work is set in $ R = 4 $. A specific study of the effects of variation of this parameter can be found in \cite{solovey04}. Infected cells of types A and B evolve differently: while the first type survives by $\tau=4$ time steps, the latter is eliminated, 
that is, it changes to \emph{dead} state in the next time step.  

The $\tau$ parameter regulates the \emph{average time} spent by the immune system to develop the \textsf{adaptative} immune response to HIV when it is at full capacity. In this way, a newly pool of infected cell will survive $\tau + 1$ weeks \emph{on average} until it starts to be is eliminated. The time interval of time $\tau$ is also associated with the peak of viral load occurring during the primary infection, which marks the effective action of the adaptive immune response. In the model proposed in \cite{santos01a}, the value $ \tau + 1$ corresponds to five weeks, which is the average characteristic value of clinically observed primary adaptive immune responses in most common viral infections, including HIV infection. The use of a fixed value for $\tau$ is a crude approximation since it does not take into account the variability of the immune response between individuals nor among different viral strains. The consideration of a density probability distribution for $\tau$ with appropriated mean and 
dispersion certainly could improve the model. However, an investigation developed in the reference \cite{figueiredo08b} shown that the variation such parameter in the interval  $ 4 \leq \tau \leq 9$ reveals that the three-stage dynamics profile remains qualitatively invariant and no substantial change occurs in the clinical latency period. Therefore, for the present purpose of describing the long-term evolution of HIV infection after the primary infection such improvement seems not to be crucial, since from the technical point of view the parameter $\tau$ only calibrates or adjust the short-time scale of the CA model.

Finally, the majority of dead cells are replaced by new CD4$^+$T cells produced by the bone marrow machinery in most of their fullness with probability $p_r$, while a fraction $(1-p_r)$ remains \emph{dead}. In the meanwhile a proportion $p_i \; p_r$ of the new CD4$^+$T cells are replaced by \emph {type-A} infected cells and $(1-p_i)\,p_r$ by healthy cells. The probability $p_i$, which is responsible for the feedback of new infected cells in the \emph{arena} mimics all mechanisms that enable the HIV to evade the immune response such as the high HIV mutation rate \cite{cuevas15} and the existence of reservoirs of \emph{latent} infected cells lodged in other extra-nodes lymphatic compartments \cite{cao18,bruner19,szu-han18}. An analysis of the robustness of the dynamics of the HIV infection with respect to $p_i$ and $p_r$ parameters points out that qualitatively the three-stage evolution remains unchanged under the variation of orders of magnitude of these quantities, only producing a temporal displacement of 
the periods associated with the peak of the primary infection and the latency period \cite{figueiredo08b}.

The value parameters were establish from experimental data following \cite{santos01a}: $p_{\tiny\textsf{HIV}}=0.05$ was chosen based on the observation that one in $10^2$ or $10^3$ T cells harbor viral DNA during the primary infection; $p_i=10^{-5}$ is due to the fact that only one in $10^4$ to $10^5$ cells in the peripheral blood of infected individual expresses viral proteins; and $p_r=0.99$ to represent the high ability of the immune system to replenish the depleted cells. See \cite{schnittman89,schnittman90}.\\

In Table \ref{table1}, the essential  parameters of the model are summarized following \cite{santos01a}.
\begin{table}[h]
\begin{tabular*}{\columnwidth}{c|l|c} \hline
 {Parameters}		& {Description} 							& {Value} 	\\ \hline\hline
 $L$          		& Lattice Linear size						& 700 		\\ \hline
 $p_{\tiny\textsf{HIV}}$    	& Initial fraction of infected cells		& 0.05  \\ \hline
 $p_r$        		& Fraction of replaced dead cell 				& 0.99 		\\ \hline  
 $p_i$        		& Fraction of infected replaced cells 			& 0.00001 \\ \hline
 $R$          		& Minimum number of infected cells type B for contact propagation	& 4 \\ \hline
 $\tau$       		& Time delay (weeks): infected cell type A  $\to$  infected 	cell type B & 4 	\\ \hline
\end{tabular*}
\caption{Parameters used in the original simulations in the CA model \cite {santos01a} to describe the HIV infection dynamics.}\label{table1}
\end{table}

\begin{figure}[ht]
\begin{center}
\includegraphics[width=0.95\columnwidth]{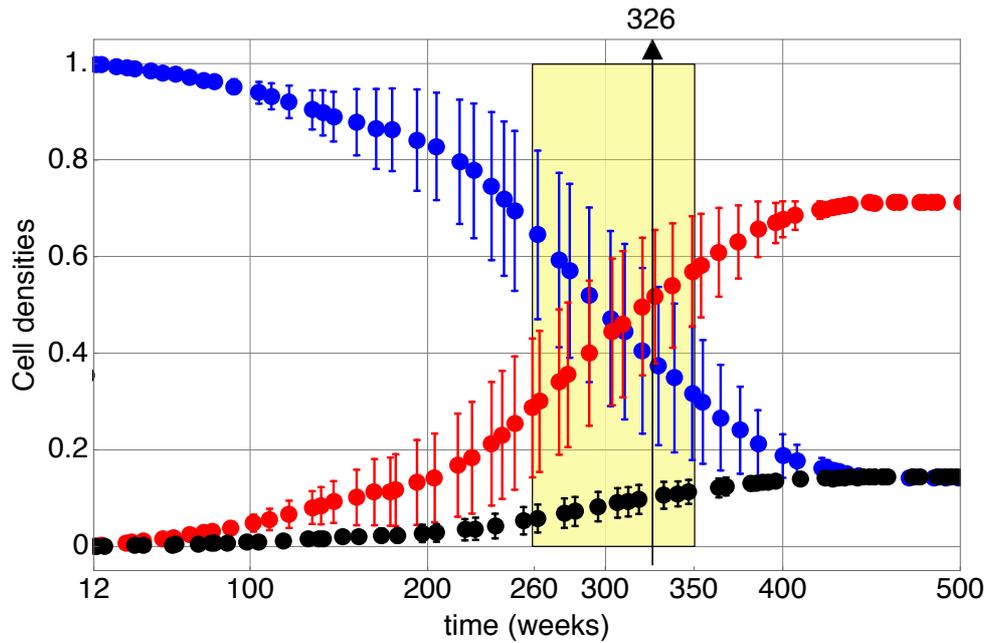}
\caption{Fractions of CD4 T$^+$ cells \emph{healthy} (blue), \emph{infected} (A e B) (red) and \emph{dead} (black) as a function of time (weeks). Note that the time scale stars at week 12 after the occurrence of the primary infection stage. The error bars indicate the standard deviation concerning the average over 100 simulations (patients). The 326-week instant and the yellow-stripe time interval (260, 350) mark relevant events in HIV dynamics explained later.}
\label{figure1}
\end{center}
\end{figure}

The dynamics of the HIV infection generated by the simulations from the CA model proposed in \cite{santos01a} can be summarized in the graphs of Figure \ref{figure1}, which illustrates the temporal evolution of the average fractions of CD4$^+$ T \emph{healthy}, 
\emph{infected}(A + B) and \emph{dead} cells just after the primary infection. The 326-week instant and the time interval (260, 350), bounded by a yellow stripe, mark essential events in HIV dynamics, which will be explicitly revealed by the three-level model later.

\subsection{Three level system}

In a recently published paper \cite{frenkel}, an energy model for a three-level system was proposed to study problems related to the Gibbs entropy.  Such model considers a system of $N$ spins, whose energy states are distributed among three energy levels. Therefore, the total number of spins $N$ with total energy $E$ can be viewed as distributed between such levels characterizing a given \emph{configuration}. Each \emph{configuration}, which may be characterized by the number of spins accommodated on each level (populations), corresponds to many microscopic states provide the spins are considered as indistinguished particles and can be permutated within each level. One of such \emph{configuration}, however, should correspond to the \emph{ground state} fixed as the \emph{zero energy} state.

The population $N_i$ and the respective energy $\mathcal E_i$ for the states $i=1,\, 2$ and 3 are related accordingly the following conservation equations for the total particle number and energy:
\begin{equation}\label{eq1}
 N_1 + N_2 + N_3 = N
\end{equation}
\begin{equation}\label{eq2}
\mathcal E_{1} N_{1}+\mathcal E_2N_2 + \mathcal E_3N_3 =E
\end{equation}

For a fixed total energy $E$ the number of possible microstates or indistinguishable configurations that meet the conditions (\ref{eq1}) e (\ref{eq2}) is given by $N!/(N_1! N_2! N_3!)$. Hence the Boltzmann's entropy $S$ of this system when $N \gg 1$ can be written, accordingly to the Stirling's approximation,  as:
\begin{equation}\label{eq3}
S = k_{B} [N\log N - N_1\log N_1 - N_2\log N_3 - N_3 \log N_3]
\end{equation}
where $k_{B}$ is the Boltzmann constant.
\\[2ex]
\subsection{Negative absolute temperatures}\label{subsec14}
The condition that characterizes the thermodynamic equilibrium is that the a variation of the entropy relative to the energy is constant and equal to the inverse of the absolute temperature,

\begin{equation}\label{eq4}
\frac{1}{T} = \frac{\partial S}{\partial E}.
\end{equation}

In the year 1956, a seminal paper of Ramsey \cite{ramsey} opened the doors to the study of interesting behaviors presented by nuclear and magnetic systems. When the entropy of a system presents a monotonically decreasing behavior relative to energy, this system would have absolute \emph{negative} temperatures, according to the equation (\ref{eq4}). These circumstances where entropy is a convex function of energy can occur in finite energy spectrum systems when the population of the state with higher energy exceeds those of the lower energy states. In these cases, the energy distribution function, characterized by the Boltzmann factor, presents an unusual behavior, that is, an exponential growth.
\begin{equation}\label{eq5}
P(E) \sim e^{E/k|T|} 
\end{equation}
with $T<0$.
Behaviors of this nature have recently been found in localized systems with finite and discrete energy spectra, such as the limited spectrum quantum systems \cite{braun,dunkel}. 

In the subsequent section, we consider a three-level model to explore the corresponding significance of negative temperature states in the process of HIV infection.

\section{The three-level system model for HIV infection}\label{sec2}

\begin{figure}[ht]
\begin{center}
\includegraphics[width=0.8\columnwidth]{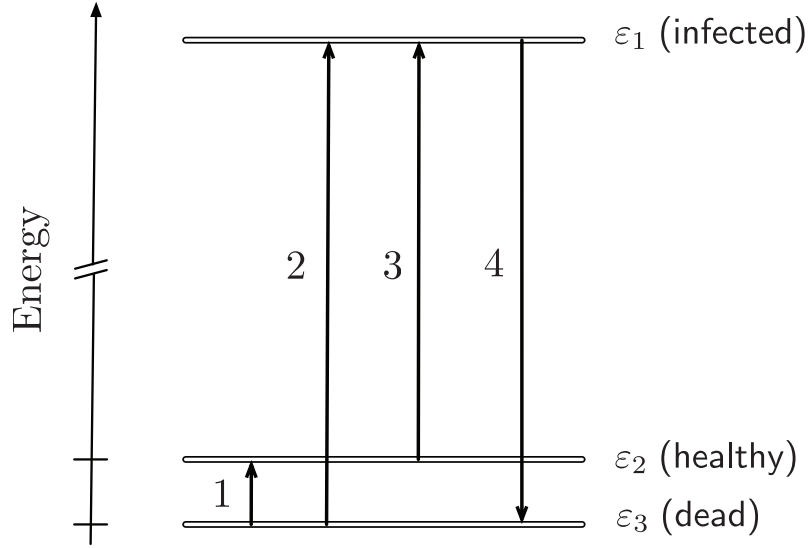}
\caption{Scheme of the energy levels and the transitions between the states of the three cell populations prescribed by the model: $\mathcal E_1$ for infected, $\mathcal E_2$ for healthy and $\mathcal E_3$ for dead cells respectively. The arrows indicate the direction of possible transitions.}
\label{figure2}
\end{center}
\end{figure}

In this Section, we design a three-level physical model to describe the dynamics of HIV infection (without treatment) analogous to that proposed in \cite{frenkel}, considering the CD4$^+$T cell population characterized by its three possible states: \textit{infected} $N_1(t)$,  \textit{healthy}  $N_2(t)$ e  \textit{dead} $N_3(t)$. The first state comprising all infected cells types A and B, the second describing the cells not reached by the viruses, and finally the third portion of cells \emph{dead}. Figure \ref{figure2} illustrates the energy and transition diagram of the proposed model.
This diagram represents the possible relative energy levels, whose values will be determined below, and the possible transitions between the three characteristic states of the infection dynamics according to the rules described above. The relative values between these energies are fixed from the populations of each state in the condition set to the threshold of the onset of AIDS, that is, when $n_1^*\sim 0.7$, $n_2^*\sim 0.2$ and $n_3^*\sim 0.1$, respectively, where $n_i=N_{i}/N$, ($i=1,\,2,$ and 3).

In terms of the variables of the HIV infection model, the equations (\ref{eq1}-\ref{eq3}) can be rewritten as:
\begin{eqnarray}
& &n_1(t) + n_2(t) + n_3(t) = 1\label{eq6}\\[2ex]
&&\mathcal E_1 n_1(t)+ \mathcal E_2 n_2(t) +\mathcal E_3 n_3(t)=\mathcal E(t) \label{eq7}\\[2ex]
& &\mathcal S(t) = - n_1(t) \log n_1(t) - n_2(t) \log n_2(t) - n_3(t) \log n_3(t) \label{eq8}
\end{eqnarray}
In the equations above $\mathcal E(t)$ and $\mathcal S(t)$ labels the energy (in units of $k_{B}T$) and the entropy (in units of $k_{B}$) per particle at the instant $t$, respectively.

The infection process is a dynamic process. In the diagram shown in the figure \ref{figure2} we indicate the possible \emph{transitions} between the three energy levels. 
The lowest energy state corresponds to that of dead cells. From this, transitions can occur to the two upper levels corresponding to the probability $P_{D \to H}$  
of the dead cells are replaced by healthy cells produced by the immune system or by infected cells, with probability $P_{D \to I }$, from infected latent cell reservoirs corresponding to transitions 1 and 2 of the figure (2), respectively. Another transition (transition 4) can happen between the infected state and the state of dead cells due to the action of the immune response after a specified mean time interval, simulated in the model by $\tau+1$ time steps. 

In our correspondence between the CA model and the 3-level energy model the  transition probabilities $P_{D \to H}$ and  $P_{D \to I}$, which represents the reposition of dead cells, are mimicked in the 3-level model by the respective Boltzmann weights, as indicated in the equations below:
\begin{equation}
P_{D \to H}= p_{ r}(1-p_{ i}) \;\propto \;e^{-(\mathcal E_2-\mathcal E_3)}\qquad \textrm{and}\qquad P_{D\to I}=p_{ r}p_{ i}\;\propto \;e^{-(\mathcal E_1-\mathcal E_3)},
\end{equation} 
where $\mathcal E_i (i=1,2,3)$ are the energies in units of $k_B T$. As can be seen in Figure \ref{figure2}, the direct transition from the infected to the healthy state is absent once this process  is prohibited by any mechanism. 

The transition probabilities $P_{D\to H}$ and $P_{D\to I}$, corresponding to transitions 1 and 2 shown in Figure 2, were calculated according to equation (9). Note that  $P_{D\to H} = p_r (1-p_i)$ is the probability that dead cells will be replaced by healthy (probability $p_r$) and non-infected cells (probability $(1-p_i)$), while $P_{D\to I}$ is the probability of dead cells being replaced by healthy cells and infected (probability $p_i$). Based on the values used for the parameters, $p_r = 0.99$ and $p_i = 10^{- 5}$, energy levels were estimated at $\varepsilon_1$ (infected) = 3.82939; $\varepsilon_2$ (healthy) = -7.68353 and $\varepsilon_3$ (dead) = -7.69359 in $k_B T $units.  Notice that the high value of $ p_r = 0.99 $ reflects one of the central assumptions of the base AC model, which assumes that bone marrow cd4T cell production machinery remains fully functioning without being affected by infection. However, the small fraction ($ 1-p_r = 0.01) $ that is not replaced at any given time step can be replaced at the next step and so on.

\section{Results and discussion}\label{sec3}
To represent the dynamics of HIV through a three-level system, we performed $N_S$ simulations of the cellular automata model proposed in \cite{santos01a} considering the distinct initial conditions but with the same values of the parameters displayed in Table (\ref{table1}). At every time step $t$ the mean values of the populations of cells $N_1(t)$,  $N_{2}(t)$ e  $N_3(t)$ are recorded, and we calculate the energy $\mathcal E(t)$ and the entropy $\mathcal S(t)$ according to equations (\ref{eq7}) and (\ref{eq8}).

\begin{figure}[hbtp]
\begin{center}
\includegraphics[width=0.95\columnwidth ]{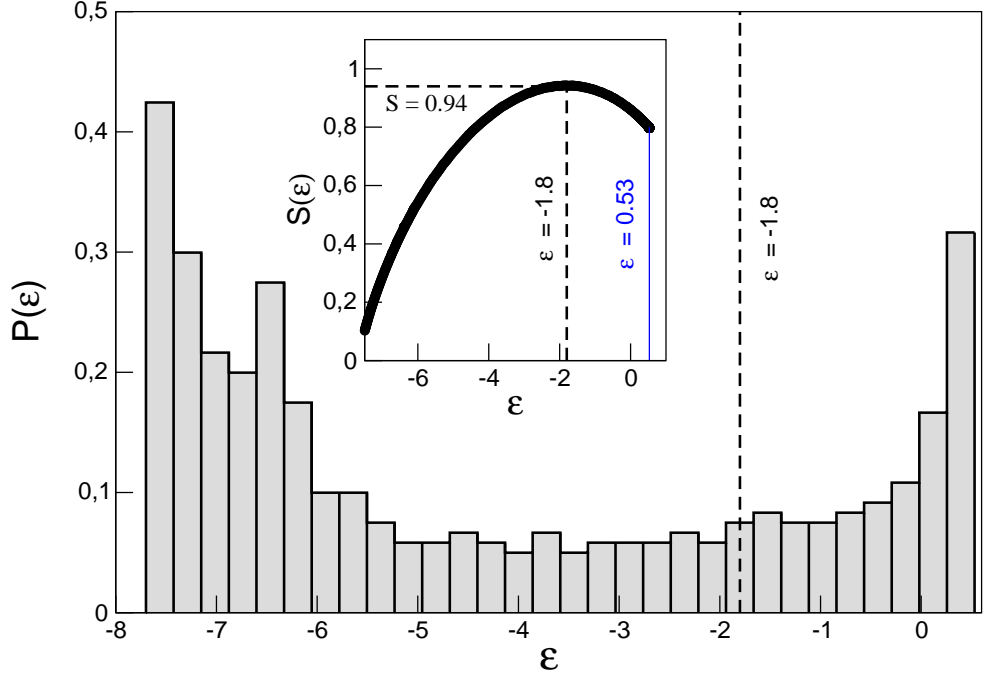}
\caption{Histogram of the energy distribution $P(\mathcal E)$ for energy states with positive ($\mathcal E < -1.8)$  and negative temperatures ($\mathcal E > -1.8)$. The doted line is guide for the eyes to separate the two regions. }
\label{figure3}
\end{center}
\end{figure}

In figure \ref{figure3} we show the histogram of the energy distribution in the interval $( -7.7,\, 0.5$), which corresponds to the dynamic process after the primary infection until the steady state, meaning from week 12 until week $\sim 450$. The region of the spectrum with negative energies $\mathcal E  \sim -1.8$ corresponds to states where the infected cell population does not yet surpass that of uninfected (healthy + dead),  which occurs on average until week 320 after the peak of the primary infection. Up to $\mathcal E \sim -5.4$, we observe a decay of the probability density function $P(\mathcal E)$, but with a secondary peak around $\sim -6.5$. This global behavior shows the highest occupation at the most negative energy levels due to the predominance of healthy cells at this stage.  In the dynamics of infection, this occurs during the clinical latency period until $\sim 231$ weeks (average). The observed peak around  $\mathcal E \sim 6.5$ corresponds to the dynamics of infection at the time 
interval between 165 and 188 weeks. At this phase of the dynamics, the probability of the emergence of compact structures becomes significant, which lead the course of the infection to the inexorable onset of AIDS as observed in references [9, 10] and discussed further below. After values of $\mathcal E \sim -5.4$ the p.d.f. $P(\mathcal E)$ remains virtually constant until energies $\sim -1.8$.

The spectrum region after $\mathcal E \sim -1.8$ presents a histogram more homogeneous with the growth of the number of states for higher energies. Positive energy values indicate the predominance of infected cells (high energies) when the unwanted development of AIDS is established with the collapse of the immune response from the week $\sim350$ (average).
The inset of Figure \ref{figure3} presents the time-parametrized plot of entropy versus energy. We observed that the maximum entropy value $\mathcal S_{\max}=0.94$ occurs for $\mathcal E_{\mathcal S_{\max}}=-1.8$.
For energy values $\mathcal E < \mathcal E_{\mathcal S_{\max}}$ the function $\mathcal S(\mathcal E)$  have a positive slope corresponding to positive temperature values.  On the other hand for $\mathcal E > \mathcal E_{\mathcal S_{\max}}$ the derivative slope is negative, hence the corresponding three-level physical model to equilibrium states with negative absolute temperatures, according to equation (\ref{eq4}).

\begin{figure}[h]
\begin{center}
\includegraphics[width=0.95\columnwidth ]{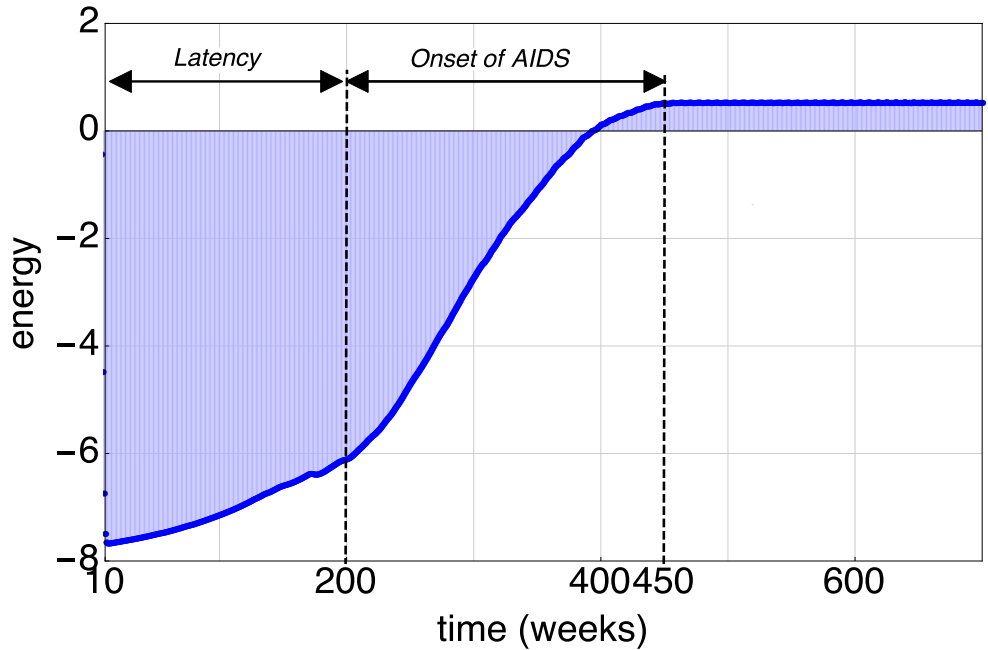}
\caption{Energy evolution from 10 to $\sim$ 200 weeks: clinical latency phase. From $\sim$200 to $\sim$450 weeks: development of AIDS.}
\label{figure4}
\end{center}
\end{figure}

\begin{figure}[h]
\begin{center}
\includegraphics[width=0.95\columnwidth ]{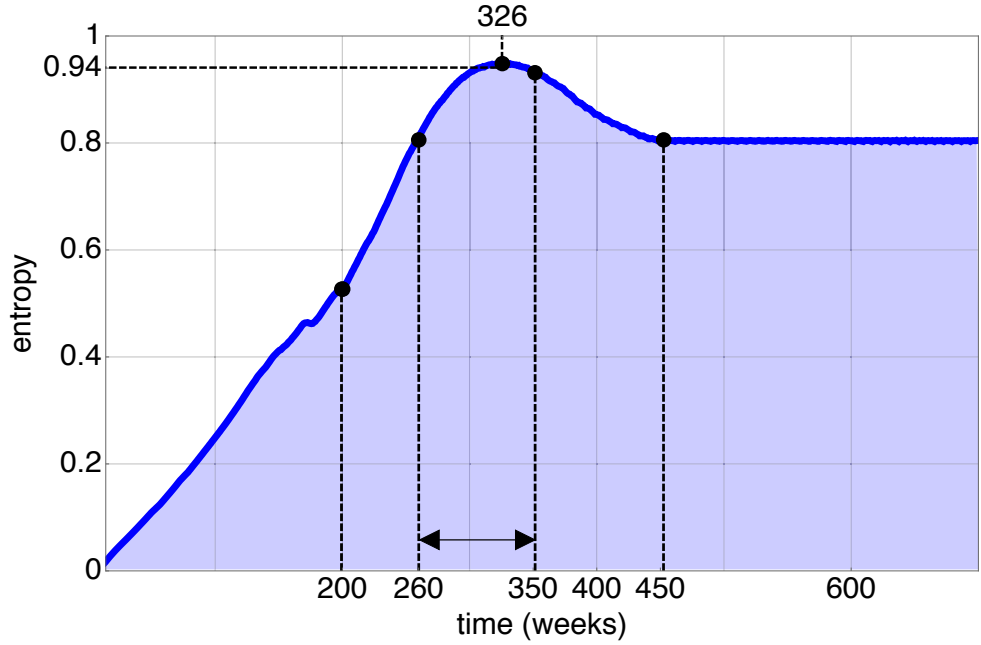}
\caption{Entropy evolution (weeks). The maximum point $\mathcal S_{\max}=0.94$ occurs in $t_{\mathcal S_{\max}}\simeq 326$ weeks. At the point ($\mathcal S\simeq 0.8, t\simeq 260$) 
the rate of change of entropy decreases until it cancels out at $t_{\mathcal S_{\max}}$. Interval $(260-350)$ weeks is the time lag for occurrence of high fluctuations  on cell concentrations. }
\label{figure5}
\end{center}
\end{figure}

We present in the figures \ref{figure4} and \ref{figure5} the time-dependent behavior of  the energy and the entropy to investigate in more detail the regions of positive and negative temperatures.

In the Figure \ref{figure4} the temporal behavior of energy is separated into three well-marked regions according to rate of energy growth:  (i) $10 < t < 200$ weeks corresponding to the clinical latency period when the density of infected cells is low and increases linearly with a time; (ii)  $200 < t < 450 $ weeks corresponding to the onset of AIDS when the density of infected cells when the rate of growth of infected cells approximately doubles and finally; (iii) $t \geq 450$ weeks when the infected cell rate reaches its maximum value and becomes stationary. In the time interval where the AIDS onset occurs, the temperature, which was previously positive, becomes negative after the  "crossing" between densities of healthy and infected cells , which happens in approximately 320 weeks, making prevalent the population of infected cells.

The Figure  \ref{figure5}  shows the behavior of the entropy of the system as a function of time.
Observe the gradual growth of the entropy until reaching its maximum value in $t\simeq 326$ weeks. Initially at an approximately constant rate during the clinical latency period (10 $\sim$ 200 weeks) followed by an increase in the growth rate up to $t \simeq 260$ weeks. Next, the growth decreases continuously to zero characterizing the instant where the temperature signal change occurs. Subsequently, the entropy decreases steadily until reaching the steady-state equilibrium value (full establishment of AIDS).
 
Comparing the graph of Figure \ref{figure5} with the behavior of the entropy against the energy, shown in the inset of Figure \ref{figure3}, we observe that the instant the entropy reaches its maximum value marks the change from the regime of positive to negative temperatures as indicated in the Figure \ref{figure3}. This value corresponds to negative but close to zero energy values.   

From the analysis of energy and entropy behaviors associated with the dynamics of HIV infection (no treatment), we conclude that the dominant state of infected cells is the one where the highest energy level is \emph{populated} with the majority of cells (particles) characterizing by negative values of absolute temperature. Since this state has most of the particles in the highest excited state hence it has the highest energy. In these circumstances, the infected state becomes the state of equilibrium of the system. 

From the temporal evolution of the entropy, we observed an almost linear monotonic growth until approximately $t\sim 200$ weeks,  when the appearance of partially ordered and compact spatial structures occur that sequester cells with predominance in the number of infected cells. Such structures originate during clinical latency and gradually grow occupying the entire network causing the development of AIDS.   Henceforth, although the mean densities, characteristic of the steady state of each cell type, a process of spatial disorder begins to occur in the compact structures evolving into configurations where spatial correlations are eliminated. In other words, with the growth of entropy in the first half of clinical latency, the system is losing information associated with the loss of dynamic correlation that happens in $t \simeq 200$ weeks. This loss of correlation can be observed through the statistical behavior of the dynamics of infection via random matrix (RM) theory \cite{ramayo17}. This RM methodology 
also shows that in the final phase of the AIDS onset the dynamic correlation is recovered reaching its maximum value described by a probability distribution function GOE (Gaussian Orthogonal Ensemble) \cite{ramayo17}.

In the Figure \ref{figure3} we found that there is a small region of energy fluctuations still with negative values but corresponding to negative temperatures. In the dynamics of infection, this region corresponds to the time interval between $\sim$ 260 and $\sim$ 326 weeks, when the ``crossing of the densities of infected and uninfected cells occurs, as mentioned above. To quantify the energy fluctuations and to establish a criterion specifying the region where absolute inversion occurs between infected and uninfected cell populations, we define the relative deviation $\sigma_\mathcal E$, as the ratio between the mean square deviation of the energy $\delta \mathcal E$ and its absolute value,
\begin{equation}\label{eq11}
\sigma_\mathcal E = \frac{\delta \mathcal E}{|\mathcal E|}.
\end{equation}

\begin{figure}[hbtp]
\begin{center}
\includegraphics[width=0.95\columnwidth ]{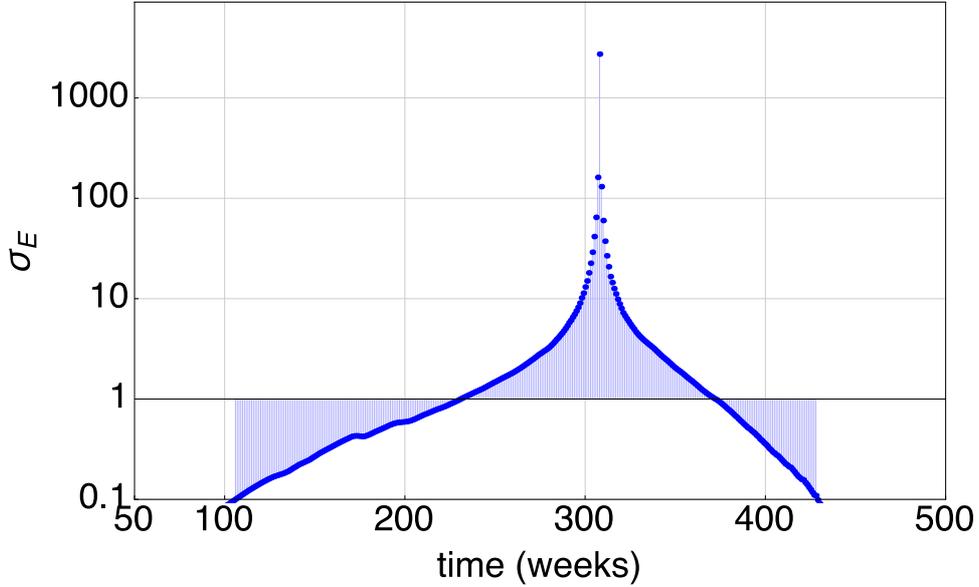}
\caption{Semi-log plot of Energy relative error dynamics, same parameters used in Figure \ref{figure1}}
\label{figure6}
\end{center}
\end{figure}

Figure \ref{figure6} illustrates the graph of this quantity as a function of time on a semi-logarithmic scale indicating a variation of three orders of magnitude. In other words, fluctuations of order $10^3$ higher than the absolute value of the energy. It appears that these fluctuations are responsible for the states in which energy and temperature are both negative. The instants of time for which the relative deviation $\sigma_\mathcal E =1$ correspond to  $t=231$ e $t=372$  weeks and mark the beginning and the end of the transition respectively. Comparing with the dynamics of infection generated by the cellular automata model for several individuals ($100$), we observe that this interval ($231-372$ weeks) encompasses the interval where error bar overlaps (r.m.s.d.)  occur of healthy and infected cell densities ($260-350$ weeks). Furthermore, the peak (divergence) where relative deviation occurs ($ \sim 309$ weeks) corresponds approximately to the point of maximum overlap where infected and healthy cell densities intersect, as shown in Figure \ref{figure1}.

\section{Concluding Remarks}

In analogy with a three-level energetic thermodynamic system, the dynamics of HIV infection described by a model of cellular automata \cite{santos01a} was analyzed revealing relevant aspects not seen before. For example, equilibrium states with positive temperatures are identified with states of negative energies $\mathcal E \leq - 1.8$ associated with states where the population of infected cells does not supplant non-infected (healthy and dead), which occurs, on average, until the 320 weeks after the primary infection. Moreover, for values of the energy spectrum $\mathcal E \leq - 5.4$ there is a marked global decay of the probability density function $P(\mathcal E)$, however, displaying a local maximum nearby $\mathcal E \simeq - 6.5$. This peak marks an essential  moment in the dynamics of CD4$^+$T cell populations  when it is observed the emergence of compact spatial structures that grow monotonically while sequestering a large number of infected cells. This situation can occur, on average, for 
the values of the 
parameters adopted in the original model \cite{santos01a}, between 165 and 180 weeks after the primary infection and is determinant to establish the clinical latency period of 200 weeks, on average. 

From the onset of compact spatial structures, the population of infected cells grows linearly at a rate well above that observed during clinical latency, accompanied by the concomitant decrease in healthy cells. In this period, known as the AIDS development, the energy grows appreciably over time, finally reaching steady state around 450 weeks. The continuous inversion in the infected cell population in this period is marked by the moment when the energy  $\mathcal E \simeq -1.8$ and the entropy reaches its value maximum, at week 326 on average. Hence, the temperature signal is reversed thereafter when the rate  of growth of the infected cells reduces reaching the steady state at week 450. The time interval between observation of the decay of $P(\varepsilon)$ ($\sim$ week 180) and the steady-state onset ($\sim$ week 450) corresponds in HIV dynamics to the interval where fluctuations in infected and healthy cell densities occur when the behavior of different individuals is put into comparison, as illustrated by the error bars in Figure 1. From the clinical point of view, the beginning of this interval marks the onset of AIDS and its concomitant development resulting in the appearance of opportunistic diseases that lead patients to death. The present study of the energy-entropic model reveals that the time interval where such fluctuations in the density of infected and healthy cells are most significant corresponds to the time interval where the mean squared deviation of energy relative to its absolute value exceeds 1 reaching a peak of three orders of magnitude in value (week $\sim$ 309), associated with crossing the average densities of healthy and infected cells, and very close to the time of occurrence of maximum entropy (week 326).

To summarize we conclude that this simple entropic-energetic analogy of the dynamics of HIV infection, presented in this study, unravel important moments in the course of the HIV infection that can not be observed and described by cellular automata models. The inclusion of antiretroviral therapy (ARV) mechanisms, as considered by the authors in references [18] is the subject of current studies. Such mechanisms of ARV therapies introduce a rapid and intense change in infected and healthy cell densities, almost instantaneous when compared to the model timescales leading the HIV dynamics probably occurring with transitions between states out of equilibrium making the correspondence with the thermodynamic energy-entropic structure conceptually tricky.




\end{document}